\RequirePackage{fix-cm}
\documentclass[twocolumn]{svjour3}          
\smartqed  
\usepackage{graphicx}
\usepackage{subfigure}
\begin{document}

\title{Pulsed Rydberg four-wave mixing with motion-induced dephasing in a thermal vapor
}


\author{
Yi-Hsin Chen         \and
Fabian Ripka         \and
Robert L\"ow  \and
Tilman Pfau
        }


\institute{
Yi-Hsin Chen \and Fabian Ripka  \and  Robert L\"ow  \and
Tilman Pfau \at
              5. Physikalisches Institut and Center for Integrated
Quantum Science and Technology, Universit\"at Stuttgart   \\
Pfaffenwaldring 57, 70569, Stuttgart, Germany \\
              \email{t.pfau@physik.uni-stuttgart.de}           
}

\date{Received: date / Accepted: date}


\maketitle

\begin{abstract}
We report on time-resolved pulsed four-wave mixing (FWM) signals in a thermal Rubidium vapor involving a Rydberg state.
We observe FWM signals with dephasing times up to 7 ns, strongly dependent on the excitation bandwidth to the Rydberg state.
The excitation to the Rydberg state is driven by a pulsed two-photon transition on ns time scales.
Combined with a third cw de-excitation laser, a strongly directional and collective emission is generated according to a combination of the phase matching effect and averaging over Doppler classes.
In contrast to a previous report~\cite{pulsedFWM} using off-resonant FWM, at a resonant FWM scheme we observe additional revivals of the signal shortly after the incident pulse has ended.
We infer that this is a revival of motion-induced constructive interference between the coherent emissions of the thermal atoms.
The resonant FWM scheme reveals a richer temporal structure of the signals, compared to similar, but off-resonant excitation schemes. A simple explanation lies in the selectivity of Doppler classes.
Our numerical simulations based on a four-level model including a whole Doppler ensemble can qualitatively describe the data.
\\
PACS 
\end{abstract}

\section{Introduction}
\label{introduction}
The field of quantum information processing by using Rydberg atoms has attracted intense interests due to the strong and long-range dipole-dipole interaction~\cite{QI_with_Rydberg}. The atomic transitions via Rydberg state lead to a blockade effect, preventing the multiexcitation within a blockade volume~\cite{RRinteraction_suppression,RRinteraction_suppression2,RRinteraction_Lukin1,RRinteraction_Lukin2,RRinteraction_Pfau}. 
It has been proposed to make use of Rydberg atoms to realize a quantum repeater~\cite{Rydberg_repeater,Rydberg_repeater2,Rydberg_repeater3}, which is an essential component for demonstrating long-distance quantum communication suffering from transmission loss~\cite{DLCZ}.
Another important requirement for a quantum network is the ability to delay or even store quantum information for a certain time. Here the bi-chromatic excitation scheme involving long-lived Rydberg states can be superior to common lambda schemes connecting different hyperfine-states by the mere frequency difference between pump and probe beams. 
Additionally, Rydberg four-wave mixing (FWM) adds nonlinearities to the medium induced from the strong interaction between Rydberg atoms, and hence the FWM process can be used for creating the non-classical light fields.
These interactions have been discussed and experimentally demonstrated for the generation of single-photon sources~~\cite{singlephotoncold,singlephotontheroy1,singlephoton_thermal} and the manipulation of quantum state~\cite{Single_transistor_Pfau,Single_transistor_Rempe,Radyber_polariton}. Therefore, the combination of FWM and blockade effects is one approach towards quantum repeaters.

Most of the relevant research is realized in Doppler-free ultracold ensembles, providing intrinsically long coherence times.
However, thermal vapors can reduce the complexity of such setups, but here the Doppler effect has to be taken into account.
By driving the transitions intensely and probing on time scales short compared to the movement of the atoms, the studies of the optical nonlinearities to single-photon level can be extended to such systems~\cite{GHzflopping,WedgeCell}.

\begin{figure} [t]
  \includegraphics[width=0.48\textwidth]{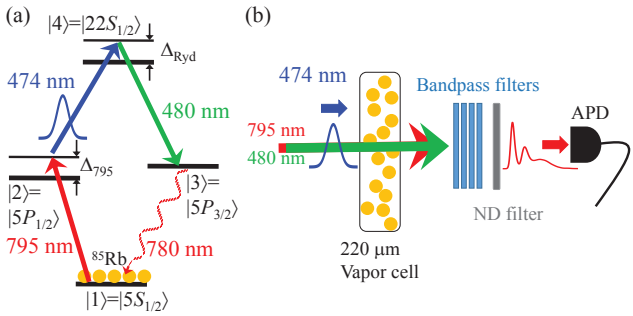}
\caption{(a) Diamond four-wave mixing scheme. A cw 795\,nm and a pulsed 474\,nm laser field drive the two-photon transition to a Rydberg state ($22S$ in our case). A 480\,nm cw-laser couples the Rydberg excitation down to the $|5P_{3/2}\rangle$ state and generates a sum-frequency of wavelength at 780\,nm due to the FWM process. (b) All laser beams are co-propagating and overlapped in a 220\,$\mu$m thick vapor cell. Four bandpass filters are inserted behind the cell for suppressing the incident light fields. Only photons generated at 780\,nm can pass through the filters. An additional ND filter is used for attenuating the light to single-photon level to provide a linear response of the APD.}
\label{fig:scheme}
\end{figure}

Coherent excitation schemes to Rydberg states have been demonstrated in cw experiments, such as electromagnetically induced transparency (EIT) \cite{WedgeCell,CW_EIT} and four-wave mixing \cite{CW_FWM,CW_Poland}, and also in pulsed schemes\cite{OhmoriUltrafast}.
Due to the Doppler effect in thermal vapors, a broad distribution of velocity classes has to be taken into account, that leads to motion-induced dephasing already on the ns time scale.
By the use of a bandwidth-limited pulsed laser system with coupling strength to the Rydberg state in the GHz regime, it is possible to observe Rabi oscillations between ground and Rydberg state on time scales below 1\,ns~\cite{GHzflopping}.
By increasing the density it is possible to add interaction effects at strengths beyond the excitation bandwidth~\cite{RRinteraction_Pfau}, similar to strongly interacting Rydberg atoms in ultracold gases \cite{ULE_cavity}.
Both are demonstrated in a ladder-type excitation scheme.
Moreover, in a pulsed FWM scheme as shown in Fig.~\ref{fig:scheme}(a), a motion-induced signal revival shortly after the incident pulse has been experimentally investigated and theoretically analyzed by considering the velocity distribution of the atoms~\cite{pulsedFWM}. The revival peak can be attributed to the constructive interference between the coherences of different velocity classes.

In this paper, we present pulsed FWM signals of durations that are long compared to the duration of the excitation pulse.
The FWM process is conducted in a resonant diamond excitation scheme, addressed from the $|5S_{1/2}\rangle$ ground state. A cw 795\,nm and a pulsed 474\,nm laser field drive the two-photon transition to the Rydberg state via an intermediate state. Another 480\,nm cw laser couples the Rydberg state to the $|5P_{3/2}\rangle$ state and converts the atomic excitation back into a light field at 780\,nm. The dynamics we studied here is beyond the frozen-gas regime, so that the atomic velocities still have a crucial influence on the temporal evolution of the generated signal. Different velocity classes exhibit different temporal phase and amplitude evolution of the signal. Therefore, the signal re-occurs after the incident pulse due to the constructive interference between different velocity classes.
This is a collective effect \cite{MandelWolfChap16}, since it originates from the interference of a group of atoms and can not be explained by single-atom behavior. As it is based on a superposition of individually acting atoms, there is no cooperativity between them. But such Rydberg-Rydberg interaction effects have already been demonstrated in thermal vapor ~\cite{RRinteraction_Pfau} and will be an essential tool in the future.

Compared to the off-resonant pulsed FWM in Ref.~\cite{pulsedFWM} ($\rm \Delta_{795}/2\pi=1$\,GHz and $\rm\Delta_{Ryd}/2\pi=200$\,MHz), the temporal structures of the FWM signals in a resonant FWM scheme exhibit more structure and are much more sensitive to the excitation bandwidth to Rydberg state. In other words, the signals are sensitive to the laser Rabi frequencies. The laser beams have spatial Gaussian shapes, so that the spatial profiles of all lasers have to be taken into account in order to achieve an agreement between experiment and theory.

\section{Experiment}
\label{sec:experiment}
We confine Rubidium atoms in a 220\,$\mu$m thick vapor cell, heated up to 140$^{\circ}$C. The atomic density is around $n = 8\,\mu$m$^{-3}$ (corresponding to $|5S_{1/2}, F=3\rangle$ of $^{85}$Rb), just low enough to neglect Rydberg-Rydberg interaction effects \cite{RRinteraction_Pfau}.
A 795\,nm and a 474\,nm laser field drive the two-photon transition to $|22S_{1/2}\rangle$ Rydberg state via the intermediate state $|5P_{1/2}, F=2\rangle$.
The 795\,nm laser and a reference 780\,nm laser are individually frequency locked with respect to the transitions of $|5S_{1/2}, F = 3\rangle$ to $|5P_{1/2}, F = 2\rangle$ and $|5S_{1/2}, F = 3\rangle$ to $|5P_{3/2}, F = 4\rangle$ (D2 transition) of $^{85}$Rb, respectively.
We then individually lock the 474\,nm and 480\,nm laser frequencies to Rydberg-EIT signals.
Note that the 780\,nm laser is only used for frequency reference, which is not sent to the vapors.
All other three incident laser beams are overlapped and pass the vapor cell in a co-propagating configuration. They are focused onto the cell with a $1/e^2$ beam diameter of around 35(5)\,$\mu$m.
The light transmission through the uncoated vapor cell is around $80\%$ for all beams.
All powers of the laser beams have been measured right in front of the cell.
The power of 795\,nm laser is varied from 5 to 200\,$\mu$W, corresponding to Rabi frequencies $\rm \Omega_{795}/2\pi$ between 55 and 345\,MHz. Because the dipole matrix element to the Rydberg state is two orders of magnitude smaller than that of the ground state transition, the pulsed 474\,nm laser field is produced by a dye amplifier with full-width-at-half-maximum of 2.3\,ns~\cite{GHzflopping} and peak power of 23\,W, implying the Rabi frequency of $\rm\Omega_{474}/2\pi = 1.1$\,GHz. The setup for the pulsed laser is similar to that in Ref.~\cite{Dye}. The power of the 480\,nm laser field is fixed to 20\,mW in all measurements, corresponding to a Rabi frequency of $\rm \Omega_{480}/2\pi=$45\,MHz.
All laser fields have the same linear polarization. 

Before sending the pulsed 474\,nm light field into the atomic ensembles, the system is initially in a steady state among a two-level system which is driven by the 795\,nm cw laser field (The 480\,nm laser field is also cw, but in steady state it has no effect on the atoms). When the pulse arrives, the intermediate and Rydberg state get coupled strongly and the transmission of 795\,nm through the atomic vapor starts to oscillates coherently \cite{GHzflopping}. At the same time, the cw 480\,nm laser drives the de-excitation to the $|5P_{3/2}\rangle$ state. The three incident light fields build up a coherence between the states $|5S_{1/2}\rangle$ and $|5P_{3/2}\rangle$, due to the four-wave mixing (energy conservation) and phase matching condition (momentum conservation).
As a consequence the generated 780\,nm field is a strongly directional emission.
In the experimental setup, four additional bandpass filters are inserted in front of an avalanche photodiode (APD) in order to suppress the stray light of the three incident fields; and a neutral-density (ND) filter is used for attenuating the generated light on few-photons level to single-photon level (in order to prevent saturation of the detector). The repetition rate of the laser pulse is 50\,Hz and all measurements are averaged up to 50,000 runs.

\section{Results}
\label{sec:results}

\begin{figure}[t]   
  \centering
  \subfigure{{\label{fig:multipeak}}
  \includegraphics[width=0.48\textwidth]{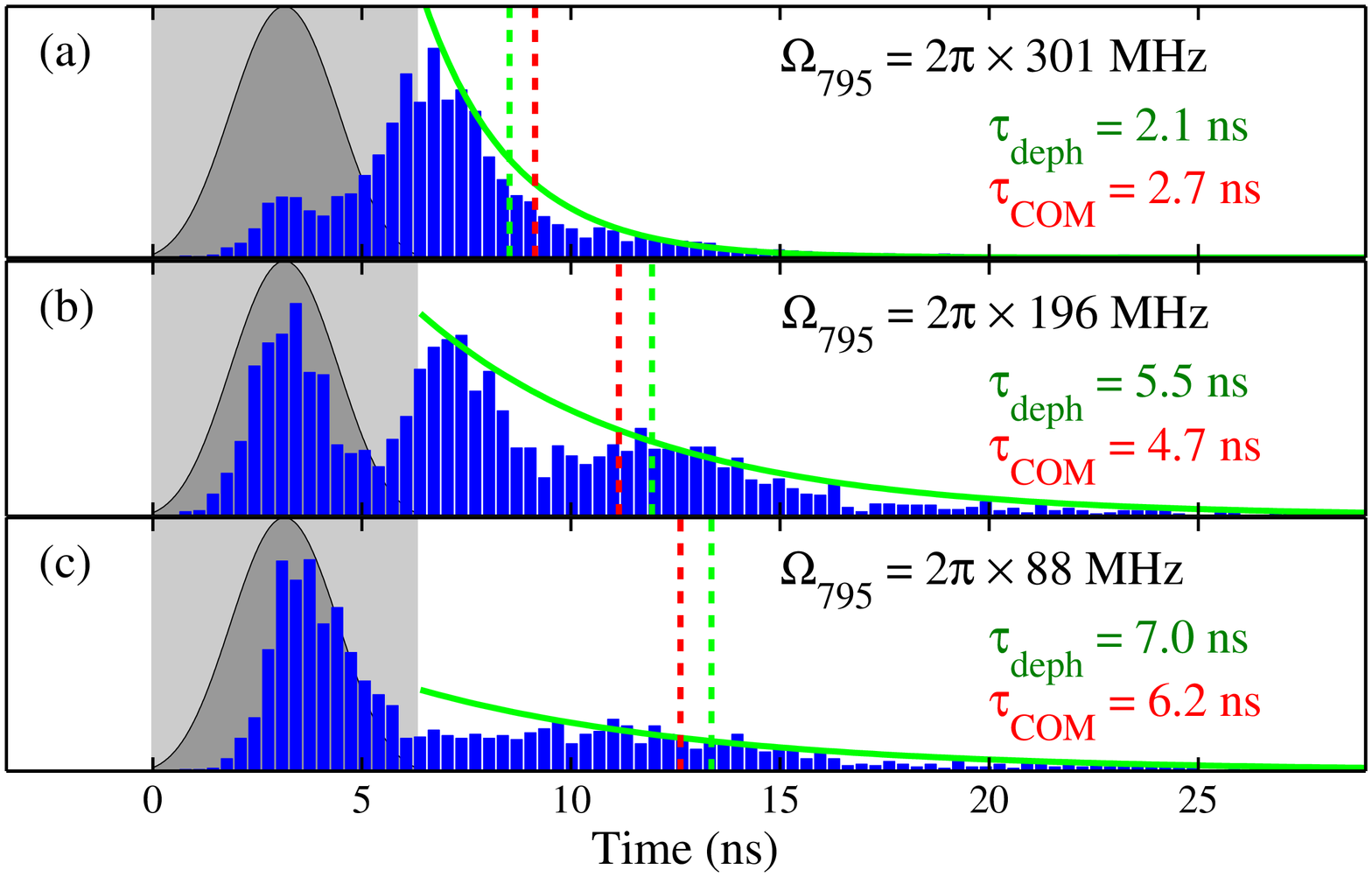}}
  \subfigure{{\label{fig:coherencetime}}
  \includegraphics[width=0.46\textwidth]{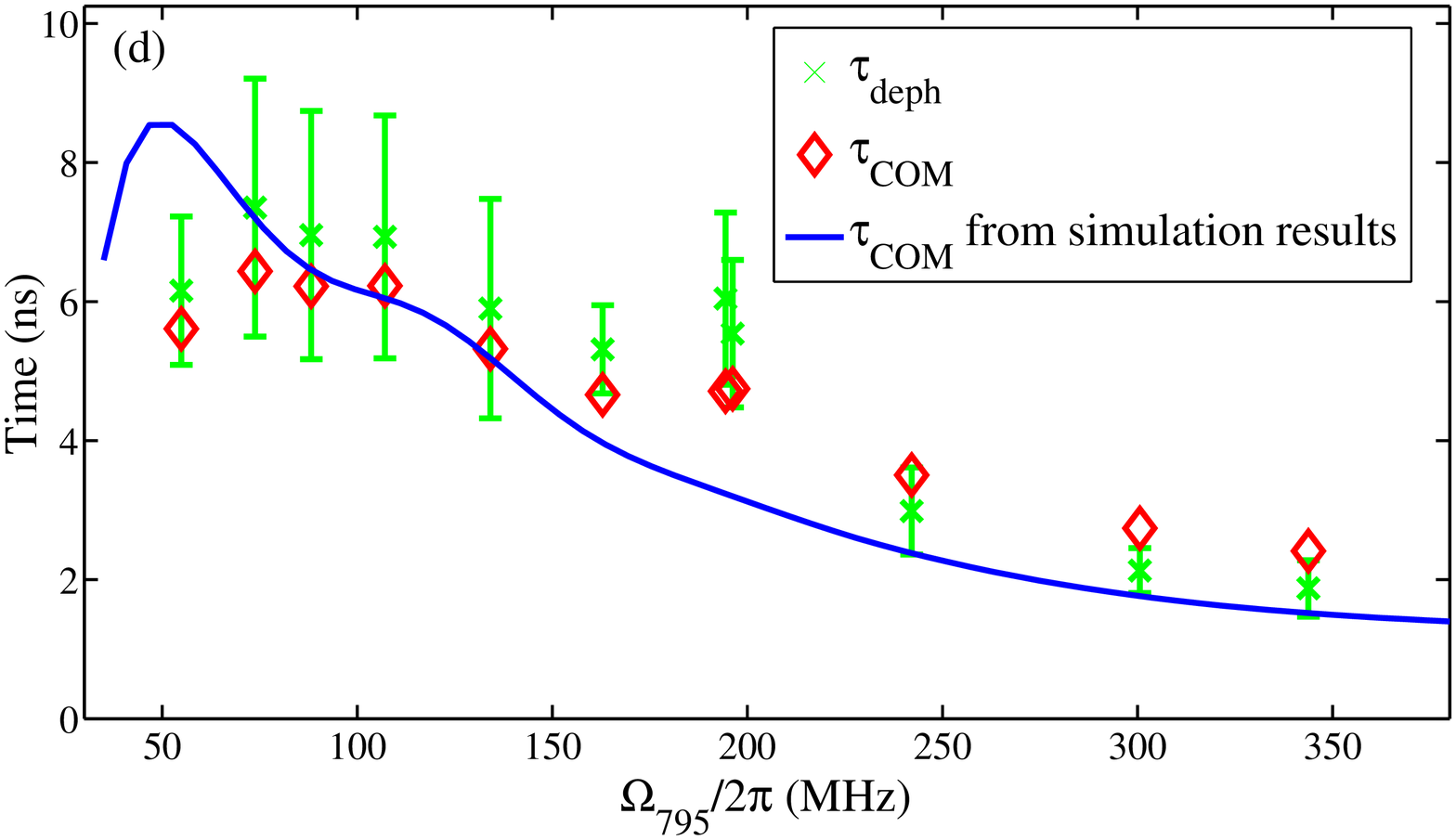}}
  \caption{(a)-(c) FWM signals (blue bars) with varied $\rm \Omega_{795}$. The dark gray area is the incident 474\,nm pulse. The input pulses and FWM signals are normalized. Green solid lines are the exponential decay functions, which represent the dephasing time of free evolution. The Rabi frequencies of the 795\,nm laser as well as the dephasing times of free evolution signals by exponential fitting ($\tau_{\rm deph}$) or center-of-mass calculation ($\tau_{\rm COM}$) are shown in the legends. The dashed lines show the timings of $\tau_{\rm deph}$ and $\tau_{\rm COM}$. (d) The extracted $\tau_{\rm deph}$ and $\tau_{\rm COM}$ versus the Rabi frequency of the 795\,nm laser. $\tau_{\rm COM}$ extracted from the theoretical predictions by considering the entire Doppler distribution and the distribution of intensities of the laser fields is shown as solid blue line. In the simulations, $\rm \Omega_{474}/2\pi=1.1$\,GHz and $\rm \Omega_{480}/2\pi=45$\,MHz.}
\label{fig:Fig2}
\end{figure}

The temporal structure of the generated FWM signals strongly depends on the individual powers for the 795\,nm laser of the resonant FWM scheme, as shown in Fig.~\ref{fig:Fig2}(a)-\ref{fig:Fig2}(c).
After the incident laser pulse, the FWM signals at weaker 795\,nm laser fields show longer dephasing times. We systematically investigate the dephasing times of the signals by varying the 795\,nm laser power and extract the dephasing times $\tau_{\rm deph}$ by applying the exponential decay functions and fitting with both the whole data and the peaks of the signals after $t = t_0 = 6.4$ ns. The averaged exponential decay time constants are 2.1, 5.5, and 7.0\,ns from Fig.~\ref{fig:Fig2}(a) to Fig.~\ref{fig:Fig2}(c) and the green solid lines are exponential functions with above-mentioned time constants. $\tau_{\rm deph}$ as a function of $\rm \Omega_{795}$ is shown in Fig.~\ref{fig:Fig2}(d). In addition, we use another convenient method to analyze the dephasing times of the complicated temporal structures by calculating the timings of center of mass ($\tau_{\rm COM}$) of the signals behind $t_0$. The timings of $\tau_{\rm deph}$ and $\tau_{\rm COM}$ are also shown in Fig.~\ref{fig:Fig2}(a)-(c) and plotted as stars and diamonds, respectively, in Fig.~\ref{fig:Fig2}(d).

The time-evolution of the atom-light system can be calculated by a theoretical model discussed in Ref.~\cite{pulsedFWM} by considering the entire Doppler distribution and the distribution of intensities of the light fields. The Gaussian velocity distribution is divided into discrete intervals with 250 velocity classes. The width of the intervals is inversely proportional to the probability of occurrence of the corresponding velocity class, i.e., a finer interval for central velocity class.
We numerically solve the Lindblad equation with the Hamilton operator in a four-level diamond configuration \cite{EIT_theory}. In the simulations, the spontaneous decay rates from intermediate states are $\rm \Gamma_{2}/2\pi=5.7$\,MHz and $\rm \Gamma_{3}/2\pi=6.0$\,MHz (from state $|2\rangle$ and $|3\rangle$ respectively); from the Rydberg state to each intermediate state they are $\rm \Gamma_{42}/2\pi=\rm \Gamma_{43}/2\pi=8$\,kHz and transient effects due to the atomic motion give rise to effective damping of 500\,kHz which both are negligible on the time scale of our experiment \cite{transient}.
For the off-resonant FWM scheme, the authors in Ref.~\cite{pulsedFWM} observe a second peak in the FWM signals.
Even without considering the intensity distribution of laser beams, the simulations can fit the measurements quite well.
However, for the resonant case, the temporal structure of the FWM signals is very sensitive to laser intensities, so that the spatial profiles of all laser fields need to be taken into account. In the calculations, we assume that the laser beams have a Gaussian profile with a $1/e^2$ beam diameter of 35\,$\rm\mu$m and the peak Rabi frequencies are derived from the laser power and the beam size. With the similar numerical calculation tools, we observe multi-revival FWM signals after the incident pulse as also revealed in the measurements. The simulation results can qualitatively fit our experimental observation. We further extract the values of $\tau_{\rm COM}$ from the numerical results and show them as a blue solid line in Fig.~\ref{fig:Fig2}(d).


\begin{figure} [t]   
  \includegraphics[width=0.48\textwidth]{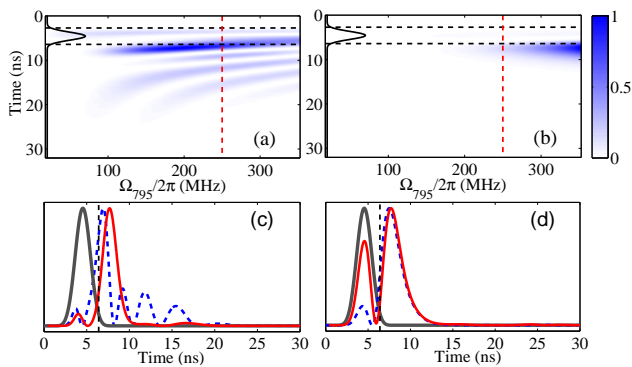}
\caption{Theoretical simulation for resonant FWM scheme (all laser detunings are zero) in (a) and that for off-resonant FWM scheme ($\rm \Delta_{795}/2\pi$ = 1\,GHz and $\rm \Delta_{Ryd}/2\pi$ = 200\,MHz) in (b). In the calculation $\rm \Omega_{474}/2\pi$ = 700\,MHz and $\rm \Omega_{480}/2\pi$ = 45\,MHz. In (c) and (d), the blue dashed lines are the temporal structures of the signals with $\rm\Omega_{795}/2\pi$ = 250\,MHz (red dashed lines in (a) and (b)). The simulation results are both normalized, and ratio of the signals in the resonant case to that in the off-resonant case is 17. The dark gray lines are the incident 474\,nm laser pulse and black dashed lines represent the timing of $t_0$. The red lines are the simulation results by considering the spatial profiles of all laser beams and using the peak Rabi frequencies as above-mentioned values.
}
\label{fig:Fig3}
\end{figure}

To understand better the origin of the rich spectroscopic structures at resonant excitation compared to off-resonant excitation, we perform numerical simulations over a larger parameter space. Therefore, we compare numerical simulations of resonant FWM with that of an off-resonant case ($\rm \Delta_{795}/2\pi = 1$\,GHz and $\rm \Delta_{Ryd}/2\pi = 200$\,MHz) in Fig.~\ref{fig:Fig3}.
From the simulation in Fig.~\ref{fig:Fig3}(b), the dynamic behaviors of the FWM signals in the off-resonant case are not very sensitive to the laser intensities, and hence,
the average values for the Rabi frequencies of the Gaussian beam distribution can be applied for a quantitative fitting.
However, in the resonant FWM scheme, the temporal structures (blue dashed line in Fig.~\ref{fig:Fig3}(c)) are much more sensitive to the laser intensities (Fig.~\ref{fig:Fig3}(a)).
By considering the spatial profiles of laser beams and summing up the contributions of each intensity distribution in the resonant case, the simulation results as shown in Fig.~\ref{fig:Fig3}(c) (red line) can qualitatively fit our experimental observation.
Therefore, in the case of resonant excitation in thermal ensembles, the spatial profiles of the laser beams play an important role of the output light behavior.

\begin{figure} [t]   
  \includegraphics[width=0.48\textwidth]{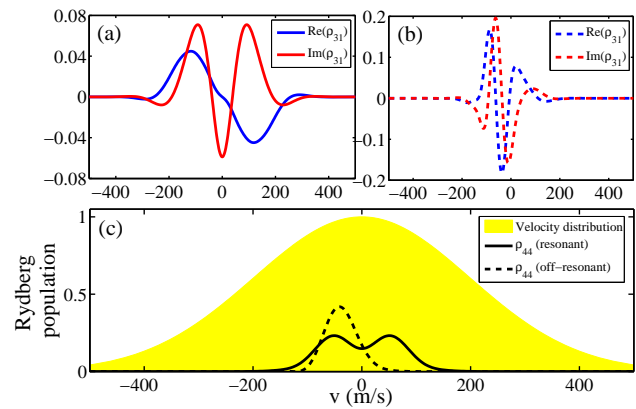}
\caption{The velocity distribution of coherence at $t_0$ in resonant and off-resonant cases in (a) and in (b), respectively.
(c) Rydberg population at $t_0$ in the resonant (solid line) and off-resonant (dashed line) cases. Yellow area is the Maxwell-Boltzmann velocity distribution. All other parameters are the same as that used in Fig.~\ref{fig:Fig3}(c) and ~\ref{fig:Fig3}(d).
}
\label{fig:Fig4}
\end{figure}

The origin of the temporal shape of the oscillatory signal lies in the interference between different velocity classes.
Due to the Doppler shift, different velocity classes experience different laser detunings, so that the corresponding coherences differ in both amplitude and phase. After the time of the incident pulse $t_0$, the atomic coherences are still evolving freely according to their Doppler detunings.
Differences in the free evolution lead to dephasing and subsequently to destructive interference of the radiation. This causes the signal to finally decrease down to zero. In the case of no driving field, the coherence of each velocity class evolves freely with the phase $\propto$ exp($-i k_{780} v t$), where $k_{780}$ is the wave-vector of the signal light and $v$ is the atomic velocity.
The ensemble-averaged coherence can be expressed by the average over the velocity distribution.
Therefore, the oscillatory behavior of the FWM signal can be referred to motion-induced revivals.
We plot the velocity distribution of coherence at $t_0$ in the resonant and off-resonant cases in Fig.~\ref{fig:Fig4}(a) and Fig.~\ref{fig:Fig4}(b), respectively.
Figure~\ref{fig:Fig4}(c) represents the Rydberg population at $t_0$ in both cases and a Gaussian velocity distribution. All other parameters are the same as that used in Fig.~\ref{fig:Fig3}(c) and ~\ref{fig:Fig3}(d). In the off-resonant case (dashed line), only a small window of velocity classes is excited to the Rydberg state due to the low excitation bandwidth.
After the pulse, as mentioned above, the coherence variation of each velocity class depends on its velocity, so that the unimodal velocity distribution leads to another maximum signal (rephasing) and the signal subsequent decreases to zero (dephasing).
On the other hand, more velocity classes are selected in the resonant scheme (solid line). The constructive interference of coherences has more combinations among the bimodal structure of velocity distribution, leading to the richer temporal structure.
Moreover, the excitation bandwidth becomes larger for a stronger 795\,nm laser field. More velocity classes involve the interference, leading to a short dephasing time. However, the dephasing time depends not only on the excitation bandwidth but also on the shape of the velocity distribution of Rydberg population. To quantitatively analyze the dephasing time, we still need the full numerical calculation.


\section{Conclusion}
\label{sec:conclusion}
We have observed velocity-induced dynamic signals in a bandwidth-limited pulsed FWM scheme via a Rydberg state. The resulting signal is an interference between the coherences of different atomic velocity classes. We systematically investigated the dephasing times as a function of intensity of the 795\,nm laser field. In contrast to off-resonant FWM, in the resonant scheme the temporal profiles of the output signals have a richer structure and are strongly sensitive to the laser power. The theoretical results by considering the entire Doppler distribution and the spatial profiles of the laser beams are consistent with the measurements.
The coherence light, with dephasing time up to 7 ns, could be preserved in the medium and retrieved by a second pulse laser, which is the mechanism for implementing a single-photon source or quantum memory \cite{singlephoton_thermal}. Our theoretical calculations and experimental finding pave the way for further studies of cooperative effects in a dense thermal vapor cell.

\begin{acknowledgements}
    The work is support by the ERC under contact No. 267100 and BMBF within Q.com-Q (Project No. 16KIS0129). Y. H. Chen acknowledges support from the Alexander von Humboldt Foundation.

\end{acknowledgements}



\begin{thebibliography}{}
%
%

\bibitem{pulsedFWM}
    B. Huber, A. K\"olle, and T. Pfau,
    ``Motion-induced signal revival in pulsed Rydberg four-wave mixing beyond the frozen-gas limit,"
    Phys. Rev. A \textbf{90} 053806 (2014).
\bibitem{QI_with_Rydberg}  
    M. Saffman, T. G.Walker, and K. M{\o}lmer,
    ``Quantum information with Rydberg atoms,"
    Rev. Mod. Phys. \textbf{82}, 2313 (2010).
\bibitem{RRinteraction_suppression}   
    D. Tong, S. M. Farooqi, J. Stanojevic, S. Krishnan, Y. P. Zhang, R. C$\rm \hat{o}$t\'{e}, E. E. Eyler, and P. L. Gould,
    ``Local Blockade of Rydberg Excitation in an Ultracold Gas,"
    Phys. Rev. Lett. \textbf{93}, 063001 (2004).
\bibitem{RRinteraction_suppression2}
    K. Singer, M. Reetz-Lamour, T. Amthor, L. G. Marcassa, and M. Weidem\"{u}ller,
    ``Suppression of Excitation and Spectral Broadening Induced by Interactions in a Cold Gas of Rydberg Atoms,"
    Phys. Rev. Lett. \textbf{93}, 163001 (2004).
\bibitem{RRinteraction_Lukin1}
    A. V. Gorshkov, J. Otterbach, M. Fleischhauer, T. Pohl, and M. D. Lukin,
    ``Photon-Photon Interactions via Rydberg Blockade,"
    Phys. Rev. Lett. \textbf{107}, 133602 (2011).
\bibitem{RRinteraction_Lukin2}
    T. Peyronel, O. Firstenberg, Q. Y. Liang, S. Hofferberth, A. V. Gorshkov, T. Pohl, M. D. Lukin, and V. Vuleti\'{c},
    ``Quantum nonlinear optics with single photons enabled by strongly interacting atoms,"
    Nature (London) \textbf{488}, 57 (2012).
\bibitem{RRinteraction_Pfau}
    T. Baluktsian, B. Huber, R. L\"ow, and T. Pfau,
    ``Evidence for Strong van der Waals Type Rydberg-Rydberg Interaction in a Thermal Vapor,"
    Phys. Rev. Lett. \textbf{110}, 123001 (2013).
\bibitem{Rydberg_repeater}
    L. H. Pedersen and K. M{\o}lmer,
    ``Few qubit atom-light interfaces with collective encoding,"
    Phys. Rev. A \textbf{79}, 012320 (2009).
\bibitem{Rydberg_repeater2}
    Y. Han, B. He, K. Heshami, C. Z. Li, and C. Simon,
    ``Quantum repeaters based on Rydberg-blockade-coupled atomic ensembles,"
    Phys. Rev. A \textbf{81}, 052311 (2010).
\bibitem{Rydberg_repeater3}
    B. Zhao, M. M\"{u}ller, K. Hammerer, and P. Zoller,
    ``Efficient quantum repeater based on deterministic Rydberg gates,"
    Phys. Rev. A \textbf{81}, 052329 (2010).
\bibitem{DLCZ}
    L. M. Duan, M. D. Lukin, J. I. Cirac, and P. Zoller,
    ``Long-distance quantum communication with atomic ensembles and linear optics,''
    Nature \textbf{414}, 413 (2001).
\bibitem{singlephotoncold}
    Y. O. Dudin and A. Kuzmich,
    ``Strongly interacting Rydberg excitations of a cold atomic gas,"
    Science \textbf{336}, 887 (2012).
\bibitem{singlephotontheroy1}
    M. Saffman and T. G. Walker,
    ``Creating single-atom and single-photon sources from entangled atomic ensembles,"
    Phys. Rev. A \textbf{66}, 065403 (2002).
\bibitem{singlephoton_thermal}
    M. M.~M\"uller, A. K\"olle, R. L\"ow, T. Pfau, T. Calarco, and S. Montangero,
    ``Room-temperature Rydberg single-photon source,"
    Phys. Rev. A \textbf{87}, 053412 (2013).
\bibitem{Single_transistor_Pfau}
    H. Gorniaczyk, C. Tresp, J. Schmidt, H. Fedder, and S. Hofferberth,
    ``Single-Photon Transistor Mediated by Interstate Rydberg Interactions,"
    Phys. Rev. Lett. \textbf{113}, 053601 (2014).
\bibitem{Single_transistor_Rempe}
    D. Tiarks, S. Baur, K. Schneider, S. D\"{u}rr, and G. Rempe,
    ``Single-Photon Transistor Using a F\"orster Resonance,"
    Phys. Rev. Lett. \textbf{113}, 053602 (2014).
\bibitem{Radyber_polariton}
    D. Maxwell, D. J. Szwer, D. Paredes-Barato, H. Busche, J. D. Pritchard, A. Gauguet, K. J. Weatherill, M. P. A. Jones, and C. S. Adams,
    ``Storage and Control of Optical Photons Using Rydberg Polaritons,"
    Phys. Rev. Lett. \textbf{110}, 103001 (2013).
\bibitem{GHzflopping}
    B. Huber, T. Baluktsian, M. Schlagm\"uller, A. K\"olle, H. K\"ubler, R. L\"ow, and T. Pfau,
    ``GHz Rabi Flopping to Rydberg States in Hot Atomic Vapor Cells,"
    Phys. Rev. Lett. \textbf{107}, 243001 (2011).
\bibitem{WedgeCell}
    H. K\"ubler, J. P. Shaffer, T. Baluktsian, R. L\"ow, and T. Pfau,
    ``Coherent excitation of Rydberg atoms in micrometre-sized atomic vapour cells,"
    Nat. Photon. \textbf{4}, 112 (2010).
\bibitem{CW_EIT}
    A. K. Mohapatra, T. R. Jackson, and C. S. Adams,
    ``Coherent Optical Detection of Highly Excited Rydberg States Using Electromagnetically Induced Transparency,"
    Phys. Rev. Lett. \textbf{98}, 113003 (2007).
\bibitem{CW_FWM}
    A. K\"olle, G. Epple, H. K\"ubler, R. L\"ow, and T. Pfau,
    ``Four-wave mixing involving Rydberg states in thermal vapor,"
    Phys. Rev. A \textbf{85}, 063821 (2012).
\bibitem{CW_Poland}
    M. Parniak and W. Wasilewski,
    ``Interference and nonlinear properties of four-wave-mixing resonances in thermal vapor: Analytical results and experimental verification,"
    Phys. Rev. A \textbf{91}, 023418 (2015).
\bibitem{OhmoriUltrafast}
    N. Takei, C. Sommer, C. Genes, G. Pupillo, H. Goto, K. Koyasu, H. Chiba, M. Weidem\"uller, K. Ohmori,
    ``Direct observation of ultrafast many-body electron dynamics in a strongly-correlated ultracold Rydberg gas,"
    arXiv:1504.03635.
\bibitem{ULE_cavity}
    R. L\"ow, H. Weimer, J. Nipper, J. B Balewski, B. Butscher, H. P. B\"uchler, and T. Pfau,
    ``An experimental and theoretical guide to strongly interacting Rydberg gases,"
    J. Phys. B: At. Mol. Opt. Phys. \textbf{45} 113001 (2012).
\bibitem{MandelWolfChap16}
    L. Mandel and E. Wolf, OPTICAL COHERENCE AND QUANTUM OPTICS, \textbf{805}, Cambridge University Press, New York, USA (1995).
\bibitem{Dye}
    A. Schwettmann, C. McGuffey, S. Chauhan, K. R. Overstreet, and J. P. Shaffer,
    ``Tunable four-pass narrow spectral bandwidth amplifier for use at $\sim$508 nm,"
    Appl. Opt. \textbf{46}, 1310 (2007).
\bibitem{EIT_theory}
    M. Fleischhauer, A. Imamoglu, and J. P. Marangos,
    ``Electromagnetically induced transparency: Optics in coherent media,"
    Rev. Mod. Phys. \textbf{77}, 633 (2005).
\bibitem{transient}
    J. Sagle, R. K. Namiotka, and J. Huennekens,
    ``Measurement and modelling of intensity dependent absorption and transit relaxation on the cesium D1 line,"
    J. Phys. B: At. Mol. Opt. Phys. \textbf{29}, 2629 (1996).

\end{thebibliography}
\end{document}